\documentstyle[12pt,epsfig]{article} 
\newlength{\dinwidth}                       
\newlength{\dinmargin}                      
\newcommand{\pmb}[1]{%
        \setbox0=\hbox{#1}%
        \kern-.02em\copy0\kern-\wd0
        \kern+.04em\copy0\kern-\wd0
        \kern-.02em\raise.0217em\box0}

\newcommand{\lsim}{
 \mathrel{\setbox0=\hbox{$<$}\raise0.6ex\copy0\kern-\wd0
 \lower0.65ex\hbox{$\sim$}}}
\newcommand{\gsim}{
 \mathrel{\setbox0=\hbox{$>$}\raise0.6ex\copy0\kern-\wd0
 \lower0.65ex\hbox{$\sim$}}}

\begin{document}

\centerline{\Large\bf Possible evidence for   color 
transparency from}
\centerline{\Large\bf dijet production with large rapidity gaps in
 $\gamma p$ scattering} 
\centerline{\Large\bf  at HERA and how to test it in $\gamma p, \gamma A$ scattering}
\vspace{0.4cm}
\centerline{\large L.Frankfurt$^{a,b}$ and 
  M.  Strikman$^{c,b}$}

\vspace{0.2cm}

\noindent
$^{a}$ School of Physics and Astronomy, 
             Tel Aviv University, Tel Aviv 69978, Israel \\ 
$^{b}$ Institute for Nuclear Physics, St. Petersburg, Russia \\
$^{c}$ Department of Physics, Pennsylvania State University, 
               University Park, PA 16802, USA

\begin{quotation}
\noindent
{\bf Abstract:}
We argue that the probability of gap survival in dijet 
production
in $\gamma p$ scattering  as measured by ZEUS may be 
due to the color
transparency phenomenon and suggest ways to test 
this hypothesis
in the future $\gamma p$ and $\gamma A$ processes.
\end{quotation}

The interaction of spatially small systems with a hadron has 
been  the subject of
discussions for a long time now (for the long and 
somewhat contradictory
history of the theoretical and experimental 
investigations of this phenomenon
see ref \cite{FMS}). One expects that 
small color singlets interact weakly if energies are not 
extremely high - color transparency (CT).
The current HERA data are in the kinematic region
 where the  coherent length  
$l_c={1/2m_Nx}$ significantly exceeds the nucleon radius. 
In this kinematic range color coherent effects should 
reveal themselves most
clearly. Here we explain a  practical idea how to search for CT 
in high $p_t$ dijet production at  HERA both in 
$\gamma p$ and $\gamma A$ collisions.
\section{Gap survival for $\gamma p$ case}

In order to  study  soft interactions which 
accompany a  hard scattering,
Bjorken \cite{Bjorken} suggested to investigate  the ratio 
of the 
 cross sections of the  high $p_t$
dijet production with a large rapidity gap (LRG)
to that of dijets without a rapidity gap:
\begin{equation}
f_{ac}={\sigma(a+c\rightarrow \left(jet(p_t)+X\right) +LRG +
\left(jet(-p_t)+Y\right))
\over \sigma(a+c\rightarrow jet(p_t)+jet(-p_t)+Z)}=
\kappa P_{LRG}
\end{equation}
Here $c$ can be a proton or a nuclear target.
To account for the difference between scales of hard and soft
processes quantify the role of soft physics 
Bjorken evaluated $f_{ac}$ as the product of 2 factors:
\begin{equation}
f_{ac}\equiv \kappa
P_{RGS}.
\end{equation}
Factor $\kappa$ 
is the probability of producing a rapidity gap in hard
subprocess, while $P_{RGS}$ characterizes probability
of gap survival due to soft interactions of constituents 
which do not participate in the hard collision.
 
Natural mechanism for the colorless hard collision is the 
exchange by 2 gluons. At 
first sight this contribution should be 0.
Really it follows from  the QCD 
factorization theorem that the 
exchange by an extra gluon between the partons 
involved in a hard collision
is canceled out for the total cross section of dijet 
production.
However for diffractive processes the presence of the LRG 
trigger in the final state destroys the  cancelation between
different terms, leading  
to the factorization theorem breaking\cite{CFS}.
In perturbative QCD 
$\kappa$, can be estimated
as  the ratio of cross sections of hard 
collisions of  partons
due to a double gluon color singlet 
exchange 
to that  due to a single gluon exchange \cite{Bjorken,MT,DT},
give  $\kappa \sim 0.15$ cf.discussion in \cite{Zeppenfeld}
which depends rather weakly on $p_t$ of the jets.
 Account for
 the leading $\alpha_s\ln x$ corrections may lead to 
a certain increase of 
$\kappa$ with the length of  rapidity gap. 
$\kappa$ is different for the hard 
collisions of partons belonging to the
different representations of $SU(3)_{color}$. 
This leads to a certain dependence of  $\kappa$ on 
the kinematics and to a weak dependence on 
a projectile.

Within the framework of conventional 
soft dynamics $P_{RGS}$ should be approximately 
independent of the 
projectile. This is because of the different geometry 
of collisions 
characteristic for soft and for hard collisions.
Hard collisions are concentrated at
small impact parameters 
which are characterized
by the average slope of the diffractive cross section:
 $a +b \rightarrow X_1+X_2$, where $X_1$, $X_2$ are diffractive states.
On the contrary, soft interactions are predominantly 
peripheral,
at impact parameters increasing with energy.
This has been established experimentally via the observation of
the diffractive cone shrinkage with increase of the energy. Thus a 
reasonable approximation is 
that
$P_{RGS}$
is determined by collisions at zero impact parameters.
Within the eikonal approximation used by Bjorken \cite{Bjorken} the 
eikonal phase at zero
impact parameters is a function of the dimensionless   ratio $\sigma_{tot}(ac)/B_{ac}$, where $B_{ac}$ is the slope
 of the differential cross
section for 
the soft
$ac$ scattering.   
 We observe that this ratio is 
practically 
the same for proton and photon projectiles.
 Here for a photon projectile we use as 
a guide the vector dominance model where $B_{\gamma c} \approx B_{\pi c}$
and $\sigma_{inel} \approx \sigma_{\pi c}$.
Hence in the eikonal approximation:
\begin{equation}
P_{RGS}(p\bar p)=P_{RGS}(\gamma p).
\end{equation}
 This projectile independence is because a  collision at central impact
parameters is almost black.

A second possible source of filling the gap between the jets 
can be radiation from the two
gluon exchange. This radiation should be a small effect since
both gluons are located at the same parameter. In this case
radiation of gluons with transverse momenta $\ll p_t$ 
is cancelled out
because such a gluon can not resolve
colorless exchange, cf.\cite{Gribov}. 
 Radiation of hard gluon is suppressed by the
smallness of the coupling constant. 
Besides, this radiation is   projectile 
independent 
since it is determined by the properties of the
 2 gluon exchange.

Very recently photoproduction events which 
have two or more jets 
have been observed 
in the $W_{\gamma p}$ range $135 <W_{\gamma p}<~280 GeV$ 
with the ZEUS detector at HERA \cite{ZEUS}.
A class of the events is 
observed with little hadronic activity between 
the jets. 
The value of $f_{\gamma p}=0.07 \pm 0.03$ is
reported based on the last bin:
 $\Delta \eta \ge 3$. 
This value is rather close to the estimates in perturbative QCD 
\cite{Bjorken,MT,DT} neglecting absorptive effects due to 
interactions of spectator partons in colliding particles,
i.e.assuming $P_{RGS}\sim 1$.
It is significantly larger that the values reported by D0 \cite{D0} 
and CDF \cite{CDF} at $\sqrt{s}$=1.8 TeV: 
$f_{p\bar p}=0.0107\pm 0.0010(stat.)^{+0.0025}_{-0.0013}(sys.)$ \cite{D0}, and 
$0.0086 \pm 0.0012$ \cite{CDF}.
The difference in the gap survival probability is another 
manifestation of the lack of factorization in  the hard processes 
when  extra constraints are imposed on the event selection, see review 
in  \cite{AFS}.

We thus conclude that the probability of gap survival seems to be
an  effective probe
of soft interactions which accompany hard interactions.
Specifics of the photon projectile is that its wave function
 contains a significant $q\bar q$ component with large transverse momenta 
where color is screened. For such configurations, CT
 would lead to significant enhancement of $P_{RGS}$. 
In the ZEUS experiment the requirement of observing two high $p_t$ jets
in the acceptance of the detector have led to an effective 
selection of jets carrying a fraction of more than 0.7 of
the photon momentum. This component of the wave function is dominated by the   
small size $q\bar q$ component of the photon wave function since
 the soft component is suppressed at least by a factor $1-z$.
Hence the larger value of $f_{\gamma p}$ observed 
in this experiments as compared to  $f_{pp}$ maybe a manifestation of CT.
In other words, kinematics of  of the ZEUS experiment
may  {\bf effectively suppress the  soft component in the parton 
wave function of photon}.  One of the ways to check this 
interpretation is to investigate the dependence of $P_{RGS}$ as a 
function of the fraction of the  photon momentum 
carried by the jet. The prediction is a significant depletion of 
$f_{\gamma p}$ when this fraction  decreases to values below 0.5.
One should also try to introduce a cut for the jet  fraction larger 
that 0.7, but to avoid kinematics when the jet from accompanying
quark would fill the gap.
This may increase the color transparency effect. 

\begin{figure}
\centerline{
\epsfig{file=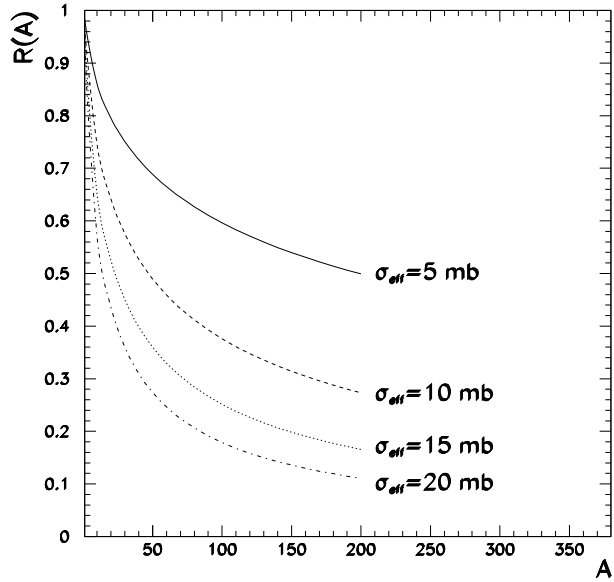,width=10.0cm,height=10.0cm}}

\vspace{-2.25cm}

\caption{$A$ dependence of the rapidity gap survival 
probability on $\sigma_{eff}$.}
\end{figure}

\section{A-dependence of gap survival}
Another way to check the color transparency interpretation of the
 ZEUS data would be to study the $A$-dependence of  $P_{RGS}$.
One can address here in {\bf a  quantitative way } the key
question of 
{\it how large is   the effective cross section for the interaction 
of the photon in the configuration which leads to the  production of events 
with rapidity gaps  between jets?} Is it close to the average  value of
 $\sigma_{eff} \sim $20 mb or maybe much smaller,  as the CT interpretation
 of the ZEUS data suggests.

Let us  define 
\begin{equation}
R(A) = { f_{\gamma A}(\Delta \eta) \over f_{\gamma p}(\Delta \eta)},
\end{equation}
for $\Delta \eta \ge 3$ where $f_p(\Delta \eta)$ flattens out.
It is easy to calculate the $A$-dependence of $R(A)$ using 
the eikonal approximation
\cite{BT}:
\begin{equation}
R(A)=\int d^2B {\tilde T}(B)\exp(-\sigma_{eff} {\tilde T}(B)).
\end{equation}
Here $ {\tilde T}(B)$ is the standard nuclear  thickness function:
$ {\tilde T}(B)=\int_{-\infty}^{\infty} d z \rho_A(\sqrt{B^2+z^2})$,
where the nuclear density $\rho_A(r)$ is normalized according to
$\int \rho_A(r) d^3r=1$. $\sigma_{eff}$ is the cross section of inelastic
 soft interaction of the  hadronic component of
 the photon wave function, excluding 
diffractive cross section.
The results of the calculation of $R(A$) are presented in Fig.1 as a function
$A$ for several values of $\sigma_{eff}$. One can see that 
measurements with nuclear targets could provide a quantitative measurement
of $\sigma_{eff}$ and hence shed a new light on the dynamics of 
strongly interacting  
color singlet object responsible for the jet events with rapidity gaps.
If one would observe $\sigma_{eff} \leq 10 mb$  this would provide
 a clear evidence  for CT in the production of dijets with LRG. 
It seems that the optimal range of the targets is $A \le 40$ since for larger
$A$,  $R(A)$  depends rather weakly on $A$.

\section{Acknowledgments}
We would like to thank A.Levy for useful comments.
This work was supported in part by U.S.Department of Energy and 
by BSF.


\begin{thebibliography}{99}
\bibitem{FMS} L. L. Frankfurt, G. A. Miller and M. Strikman, Ann. Rev. 
of Nucl.
and Particle Phys.~44~(1994)~501. 
%
%
\bibitem{ZEUS}ZEUS Collaboration,  M.Derrick et al, Phys.Lett.B369(1996)55.
 %
%
\bibitem{Bjorken}J.D.Bjorken, Phys.Rev. D47 (1992)101.
%
%
\bibitem{MT}A.H.Mueller and W.-K.Tang, Phys.Lett.B284(1992)123.
%
%
\bibitem{DT}V.Del Duca and W.-K.Tang, Phys.Lett.B312(1993)225.
%
%
\bibitem{Zeppenfeld}D.Zeppenfeld,MADPH-95-933 (1995)
%
%
\bibitem{CFS}J. C. Collins, L. L. Frankfurt and M. Strikman, Phys. 
Lett B307~(1993)~161.
%
%
\bibitem{Gribov} V.Gribov,Yad.Fiz.5(1967) 399.
%
%
%
\bibitem{D0}D0 Collaboration, S.Abachi et al, Phys.Rev.Lett.72(1994)2332;
FERMILAB-PUB-95-302-E(1995).
%
%
\bibitem{CDF}CDF Collaboration, S.Abe et al,Phys.Rev.Lett. 74(1995)855.
%
%
%
\bibitem{AFS} H.Abramowicz,  L.Frankfurt and M.Strikman, DESY-95-047; 
 SLAC Summer Inst.1994:539-574.
%
%
\bibitem{BT}L. Bertocchi and D. Trelleani, J. Phys. {\bf G3}, 147 (1977).
\end{thebibliography}
\end{document}